\documentclass[prl,superscriptaddress,twocolumn]{revtex4-1}
\usepackage{graphicx}
\usepackage{amsmath}
\usepackage{xcolor}

\begin{document}


\title{Magnetic domain wall dynamics in the precessional regime: influence of the Dzyaloshinskii-Moriya interaction }

\author{Jose Pe\~{n}a Garcia}
\affiliation{Univ.~Grenoble Alpes, CNRS, Institut N\'eel, Grenoble, France}
\author{Aymen Fassatoui}
\affiliation{Univ.~Grenoble Alpes, CNRS, Institut N\'eel, Grenoble, France}
\author{Marlio Bonfim}
\affiliation{Dep. de Engenharia Elétrica, 
Universidade Federal do Parana, Curitiba, Brasil}
\author{Jan Vogel}
\affiliation{Univ.~Grenoble Alpes, CNRS, Institut N\'eel, Grenoble, France}
\author{Andr\'{e} Thiaville}
\affiliation{Laboratoire de Physique des Solides, Universit\'{e} Paris-Saclay, CNRS, Orsay, France}
\author{Stefania Pizzini}
\affiliation{Univ.~Grenoble Alpes, CNRS, Institut N\'eel, Grenoble, France}
\email[]{stefania.pizzini@neel.cnrs.fr}


\date{\today}

\begin{abstract}
The domain wall dynamics driven by an out of plane magnetic field was measured for a series of  magnetic trilayers with different strengths of the interfacial Dzyaloshinskii-Moriya interaction (DMI). The features of the field-driven domain wall velocity curves strongly depend on the ratio of the field $H_{D}$ stabilizing chiral N\'{e}el walls to the demagnetizing field within the domain wall $H_{DW}$.  The measured Walker velocity, which in systems with large DMI is maintained after the Walker field, giving rise to a velocity plateau up to the Slonczewski field $H_S$, can be related to the DMI strength. Yet, when  $H_{D}$ and  $H_{DW}$ have comparable values, a careful analysis needs to be done in order to evaluate the impact of the DMI on the domain wall velocity. By means of a one-dimensional model and 2D simulations, we extend this method and we clarify the interpretation of the experimental curves measured for samples where $H_{D}$ and $H_{DW}$ are comparable.

\end{abstract}

\maketitle

Magnetic thin films deposited on a heavy metal with large spin-orbit coupling have been much studied since the discovery that the interfacial Dzyaloshinskii-Moriya interaction (DMI) \cite{Dzyaloshinskii1957,Moriya1960} can favour the stabilisation of chiral textures such as chiral N\'{e}el domain walls (DWs)\cite{Thiaville2012} and magnetic skyrmions \cite{Fert2017}.
The dynamics of  N\'{e}el  domain walls in the presence of DMI is strongly modified with respect to that of a Bloch wall, as first shown by Thiaville \textit{et al.} in 2012 \cite{Thiaville2012} within a 1D model formalism. One of the most appealing consequences is that the Walker field $\mu_{0}H_{W}$ - above which the magnetization within the DW starts precessing leading to a decrease of its speed  -  is proportional to the DMI strength.  The field-driven DW velocity at the Walker field scales with the ratio $D/M_{s}$, where $D$ is the DMI strength and $M_{s}$ the spontaneous magnetization of the thin film. Therefore, very large DW speeds have been measured in ferromagnetic samples for large DMI or in  ferrimagnetic films close to the compensation temperature  \cite{Pham2016,Kim2016,Chaves2019}. Furthermore, in the presence of a large $D/M_{s}$ ratio, the field-driven DW velocity was not observed to drop for fields larger than the Walker field, the DWs continuing to move   with the Walker velocity for fields well above it, giving rise to a velocity plateau \cite{Yoshimura2015,Pham2016,Krizakova2019}. In a recent paper we showed, experimentally and theoretically,  that the extension of the velocity plateau, starting at the Walker field and ending at the so-called Slonczewski field \cite{Krizakova2019}, is also proportional to $D/M_{s}$,  so that the DW velocity can keep high values up to large magnetic fields for systems with large DMI and/or small $M_{s}$ \cite{Krizakova2019}. This effect is observed in wide strips or in continuous films, where above the Walker field the DWs cannot be considered as 1D objects since vertical Bloch lines develop within them due to the precession of their magnetization. 

Measuring the plateau DW velocity after the Walker field can thus provide a valuable method to obtain the strength of the DMI, as this depends only on the knowledge of the spontaneous magnetization $M_{s}$. In this work, we study the application of this method  to a series of samples having different strengths of the DMI and different depinning fields. We show that, in samples with weak DMI and large depinning fields, care should be taken to associate the apparent saturation of the DW velocity at high fields to the presence of 2D effects related to the presence of DMI.  

For this purpose, we will discuss the results of  field-driven domain wall speed measurements obtained for Pt/Co/M (M=Pt, Ta, Au) trilayer samples with perpendicular magnetic anisotropy (PMA) and different values of the effective DMI strength, related to the different nature of the Co/M interface. These results are compared with those obtained for a Pt/Gd$_{77}$Co$_{23}$/Ta stack studied in Ref. \cite{Krizakova2019}. The Pt/Co interface is the prototypical interface hosting a large DMI.  Theoretical \cite{Yang2015,Yang2016,Freimuth2014} and experimental  studies \cite{Pizzini2014,Belmeguenai2015,Cho2015,Kim2016}  carried out on polycrystalline samples agree on the fact that such an interface is the source of strong DMI  with anticlockwise rotation of the magnetic moments (left-handed chirality). For a fixed Co thickness, the effective DMI can be strengthened if the Co layer is capped with an oxide, as proposed theoretically \cite{Boulle2016,Belabbes2017,Yang2016} and demonstrated experimentally for Pt/Co/AlOx  and Pt/Co/GdOx \cite{Chaves2019} or for Pt/Co/MgO \cite{Boulle2016}. For most of  the Pt/Co/M trilayers (M=Ta, Ir, Ta, Ru, Cu...) reported in the literature,  the effective \emph{interfacial} DMI ($D_s$=$Dt_{Co}$, in pJ/m) appears to be weaker than that obtained for Pt/Co/oxide trilayers  \cite{Hrabec2014,Ajejas2017,Kim2018,Shahbazi2018,Belmeguenai2019}; this indicates that the studied Co/M interfaces contribute with a DMI opposite to that of the Pt/Co interface. It is then possible to tune the DMI strength and the DW velocity by engineering the proper Pt/Co/M trilayer. For this work, we have chosen Pt/Co/Pt, Pt/Co/Ta/Pt, Pt/Co/Au and Pt/Gd$_{77}$Co$_{23}$/Ta, as model systems with different effective DMI strengths. We show that the field-driven velocity curves for DWs within the cobalt layer present very different features, and we compare the results with the predictions of 1D model and 2D micromagnetic simulations. As an extension of our previous work, we  derive the exact expression of the Slonczewski field in the case where the DMI field and the DW demagnetizing field are similar and where the approximation previously used is shown to be not accurate.

\begin{table*}
\caption{ Unit surface magnetization $M_{s}t$, in-plane anisotropy field $\mu_{0}$H$_{K}$, experimental DMI field $\mu_{0}H_{DMI}$, domain wall width parameter $\Delta$, effective DMI strength extracted  from the DMI field ($D_{s}^{eff,H}$), ratio $D/M_{s}$ (experimental values are indicated with a *). The Pt/Gd$_{77}$Co$_{23}$/Ta sample has been described and studied in Krizakova et al. \cite{Krizakova2019}.   }
    \begin{tabular}{p{4cm}  p{1.5cm}    p{1.2cm}   p{1.5cm}  p{1.5cm}  p{2cm}  p{2cm} } \hline
 Sample            &  $M_{s}t^{*}$ & $\mu_{0}H_{K}^{*}$ &       $\mu_{0}H_{DMI}^{*}$ & $\Delta$ & $D_{s}^{eff,H}$ & $|D/M_{s}|$ \\ \hline
                      & [mA]      & [T]                          &        [mT] &   [nm]          & [pJ/m]   &    [nJ/(A m)]           \\ \hline
 \\

    Pt/Co(0.5)/Pt         &0.65        & 1.09                            & 16        &    4.75            & -0.05   &    0.08              \\

 Pt/Co(0.8)/Ta(0.16)/Pt&  1.04       & 1.03                             & 39         &     4.89          & -0.20$\pm$0.03   &   0.20    \\

 Pt/Co(0.8)/Ta(0.32)/Pt&  0.98       & 1.37                             & 77          &      4.37        & -0.33$\pm$0.05   &  0.34  \\

 Pt/Co(0.8)/Au         &  1.1       & 1.04                              & 198         &     4.73          & -1.03$\pm$ 0.1  &  0.94  \\

Pt/Gd$_{77}$Co$_{23}$(4)/Ta           &  0.92   & 0.69                      & 90               &  9.4   & -0.78$\pm$ 0.1  & 0.85  \\

\hline

    \end{tabular}

\end{table*}

\section{Sample growth and characterisation}

Pt(4)/Co(0.5)/Pt(2), Pt(4)/Co(0.8)/Ta(t)/Pt(2)   (t=0.16~nm and 0.32~nm for the Ta dusting layer), Pt(4)/Co(0.8)/Au(4) and Pt(4)/Gd$_{77}$Co$_{23}$(4)/Ta(2) (thicknesses in nm) were prepared by magnetron sputtering on Si/SiO$_{2}$ substrates covered with a Ta(3) buffer layer. The Gd$_{77}$Co$_{23}$ layer was  prepared by co-sputtering of Gd and Co targets \cite{Krizakova2019}. All the samples exhibit out-of-plane magnetization giving rise to square hysteresis loops. Due to the anti-parallel alignment of the Co and Gd magnetic moments and the vicinity of the compensation temperature, the magnetization of the Ta(3)/Pt(4)/Gd$_{77}$Co$_{23}$(4)/Ta(2) is much lower than that of the other samples.  More details on sample preparation and characterization are given in the Supplementary Information file \cite{SupplMat}.
The spontaneous magnetization and the effective magnetic anisotropy were measured by VSM-SQUID. The magnetic parameters are reported in Table I. 

Domain wall velocity measurements were carried out using polar  magneto-optical Kerr microscopy.
The domain wall speeds were measured as a function of the out-of-plane magnetic field intensity, up to $B_{z}$=300~mT, using microcoils associated to a pulse current generator providing down to 30~ns long magnetic pulses with a rise/fall time of $\approx$ 5~ns  (see Supplementary Information for details \cite{SupplMat}). 
The film magnetization was first saturated in the out-of-plane direction. An opposite  magnetic field pulse $B_{z}$ was then applied to nucleate one or several reverse domains. The DW velocity was deduced from the expansion of the initial bubble domain, after the application of further magnetic field pulses.

\begin{figure*}
  \begin{center}
\includegraphics[width=18cm]{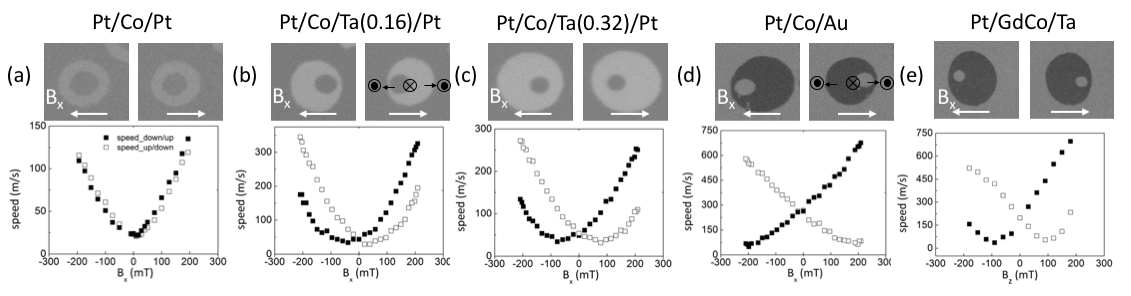}
  \end{center}
\caption{\label{fig:Figure1}\textbf{Domain wall speed versus B$_{x}$ field}: (Top): Differential Kerr images representing the expansion of a bubble magnetic domain driven by  an out-of-plane magnetic field pulse in the presence of a continuous in-plane magnetic field (white arrows); (Bottom): up/down (empty symbols) and down/up (black symbols) domain wall velocity versus in-plane magnetic field $B_{x}$ driven by an out-of-plane magnetic field $B_{z}$; (a) Pt/Co/Pt: images taken with $B_{x}=\pm$25~mT and curves with $B_{z}$=250~mT; (b) Pt/Co/Ta(0.16)/Pt: images taken with $B_{x}=\pm$125~mT and curves with $B_{z}$=240~mT; (c) Pt/Co/Ta(0.32)/Pt: images taken with $B_{x}=\pm$ 190~mT and curves with $B_{z}$=340~mT; (d) Pt/Co/Au: images taken  with $B_{x}=\pm$190~mT and curves with $B_{z}$=155~mT; (e) Pt/Gd$_{77}$Co$_{23}$/Ta: images taken  with $B_{x}=\pm$100~mT and curves with $B_{z}$=125~mT.}
\end{figure*}

\section{Determination of the interfacial DMI strength from H$_{DMI}$ field }

The sign and the strength of the DMI  can be  obtained by measuring the expansion of  bubble domains driven by an out-of-plane magnetic field pulse, in the presence of a constant in-plane magnetic field $B_{x}$ parallel to the DW normal \cite{Je2013,Hrabec2014,Vanatka2015,Pham2016}. In systems with DMI and chiral N\'{e}el walls, the DW propagation is asymmetric in the direction of $B_{x}$ and the DW speed is larger/smaller for DWs with magnetization parallel/antiparallel to the in-plane field. Differential Kerr images illustrating the asymmetric expansion of the magnetic domains are shown in Fig. \ref{fig:Figure1}. These allow us to establish  without ambiguity the left handed chirality of the N\'{e}el domain walls in all the samples (see the sketch of the magnetic moment direction within the DWs in Figure 1.)


In the presence of DMI, the DW speed reaches a minimum  when the applied in-plane field $B_{x}$ compensates the $\mu_{0}H_{DMI}$ field that stabilizes the N\'{e}el walls. 
In the thermally activated regime this condition corresponds to the field where the domain wall energy is maximum \cite{Je2013}.  Here however, we chose to work with large magnetic fields above the Walker field, where the DWs move in the precessional regime (see values in the caption of Fig. \ref{fig:Figure1}). According to our previous work this gives rise to a more reliable measurement of the DMI field \cite{Vanatka2015,Pham2016}. In  Appendix B we demonstrate that in this regime the DW velocity minimum is indeed obtained when the in-plane field compensates the DMI-induced effective field at the domain wall. Physically, this corresponds to the condition where the DW magnetization is allowed to precess as much as possible. In the Supplementary Information file \cite{SupplMat}, this is also confirmed by 2D micromagnetic simulations.    

From the in-plane field for which the DW velocity is minimum we can then deduce the effective DMI energy constant $D$ (in mJ/m$^{2}$) or the interfacial DMI constant $D_{s}$ (in pJ/m) since:
\begin{equation}
\label{eq:HDMI}
  \mu_{0}H_{DMI} = \frac{D}{M_{s}\Delta} = \frac{D_{s}}{M_{s}t \Delta }\
  \end{equation}
where $\Delta = \sqrt{A/K_{eff}}$, $A$ is the exchange stiffness (in pJ/m), $K_{eff}$ is the effective anisotropy energy (in kJ/m$^3$) and $t$ is the magnetic layer thickness.
The measured DMI strengths correspond to their effective value \textit{i.e.} to the sum of the values at the two  interfaces of the cobalt layer, $D^{eff} = D^{top} + D^{bottom}$.

The main drawback of this method, no matter the strength of the $B_z$ field, is that it requires the knowledge of the value of the exchange stiffness $A$, whose measurement is not  trivial for ultrathin films. In one of our previous works  on Pt/Co/oxide samples grown under the same conditions \cite{Chaves2019},  the best agreement between the DMI value obtained from the $\mu_0 H_{DMI}$ field and Brillouin light scattering measurements  was obtained  using $A$=16~pJ/m. On the other hand, for the Pt/Gd$_{77}$Co$_{23}$/Ta stack, the best agreement between the DMI value obtained from the $\mu_0H_{DMI}$ field and from the Walker velocity, was obtained using $A$=7~pJ/m \cite{Chaves-PhD}. These are the values used in this work.  The DW speed versus $B_{x}$ field curves are shown in Fig. \ref{fig:Figure1} for the Pt/Co/M and Pt/Gd$_{77}$Co$_{23}$/Ta samples. The $\mu_{0}H_{DMI}$ fields and the  interfacial DMI constants $D_{s}$ obtained using Eq.\ref{eq:HDMI} are reported in Table I.

The DMI is extremely small in Pt/Co/Pt, but nevertheless sufficient to favour a N\'{e}el component of the DW magnetization, giving rise to a small shift of the velocity curves for up/down and down/up domain walls ($\mu_{0}H_{DMI}$=16~mT). The presence of uncompensated DMI at the Pt/Co and Co/Pt interfaces (where we may expect same DMI strengths with opposite sign) points to the slight difference of the two interface structures, a feature often observed, see e.g. Ref.~\cite{Je2013}.

The shift of the velocity curves for up/down and down/up domain walls is larger when a dusting layer of Ta is inserted between Co and the top Pt layer, indicating the increase of the DMI strength. The interfacial DMI constants extracted from the $\mu_0H_{DMI}$ fields are   $D_{s}\approx$-0.2~pJ/m   for t$_{Ta}$=0.16~nm and  $D_{s}\approx$-0.33~pJ/m for t$_{Ta}$=0.32~nm. Referring for simplicity to the Fert and Levy  3-site indirect exchange mechanism \cite{Levy1980}, the increase of the DMI strength in the presence of the Ta dusting layer suggests that the DMI at the top Co interface decreases when a Co-Pt-Co triangle of atoms is substituted by a Co-Ta-Co  triangle at the interface. This can then explain the increase of the effective DMI and its variation with the Ta layer thickness.
If we assume that the interfacial DMI at the bottom Pt/Co interface is around $D_{s}\approx$-1.3~pJ/m, as we found for our previous samples grown in the same conditions \cite{Chaves2019}, the contribution of the top Co interface in Pt/Co/Ta(0.32)/Pt is expected to be  $D_{s}\approx$+1~pJ/m, to be compared with $D_{s}\approx$+1.2~pJ/m for the top interface in Pt/Co/Pt.  Although we are unable to quantify the ratio of Pt-Co and Ta-Co bonds at the top Co interface, this result infers a strong decrease of the DMI of the Co/Ta interface, with respect to the Co/Pt. This is in agreement with the results recently reported by Park et al. \cite{Park2018} for Pt/Co/Ta trilayers.

In Pt/Co/Au, the $\mu_{0}H_{DMI}$ field is much larger than in the previous samples ($\approx$200~mT) and corresponds to a DMI constant of $D_{s}\approx $ -1.03~pJ/m, somehow larger than those quoted  in Refs.~\cite{Park2018,Hrabec2019}, in which the layers exhibited a better crystallinity than our textured poly-crystalline samples.  With the arguments developed before, we expect that  $D_{s} \approx $+0.3~pJ/m  at the Co/Au interface. The lower values of $D_{s}$ with respect to previous works points out that the DMI strength is intimately related  the morphology of the interfaces.  

For comparison, the relatively small $\mu_{0}H_{DMI}$ field (90~mT) obtained for Pt/Gd$_{77}$Co$_{23}$/Ta, corresponds to $D_{s}\approx$-0.8~pJ/m, which is much larger than that obtained for Pt/Co/Ta(0.32)/Pt ($D_{s}\approx$-0.33~pJ/m) for a similar  $\mu_{0}H_{DMI}$. This is mainly due to the dependence of $\mu_0 H_{DMI}$ field  on the DW parameter $\Delta$ (Eq.\ref{eq:HDMI}) that is much larger in Gd$_{77}$Co$_{23}$ due to the smaller effective anisotropy (Table I).

\begin{table*}[t]

\caption{Results of 1D model of domain wall dynamics for the multilayer samples: the ratio $\delta=H_{D}/H_{DW}$ is computed from Table I; $\mu_{0}H_{W}^{(a)}=\alpha \mu_0H_{D}$ is the approximated Walker field valid when $\delta \gg 1$;  $\mu_{0}H_{W}^{(b)}$ is the exact Walker field from Eq. 4; $\mu_{0}H_{S}^{(a)}$ is the approximated  Slonczewski field using Eq. 7 with $\mu_{0}H_{W}^{(a)}$; $\mu_{0}H_{S}^{(b)}$ is obtained from Eq. 7 using $\mu_{0}H_{W}^{(b)}$ ; $\mu_{0}H_{S}^{(c)}$  is the exact Slonczewski field using a semi-analytical approach (Appendix A); $v_{W}$ is the velocity calculated at the exact $\mu_0H_{W}$ field, using Eq. 5; $v_{sat}^{*}$ is the experimental DW velocity at saturation (plateau velocity); $D_{s}^{vsat}$ is the interfacial DMI obtained from $v_{sat}$ using Eq. \ref{eq:vW}. The values in brackets are the ones extracted from the measured $\mu_0H_{DMI}$ field.   }

\begin{tabular}{p{3.5cm}  p{1.2cm}             p{1.4cm}              p{1.4cm}                 p{1.4cm}              p{1.4cm}                p{1.4cm}       p{1.0cm}  p{1.3cm}          p{2.2cm} } \hline

    Sample       &   $\delta $     &  $\mu_{0}H_{W}^{(a)}$ & $\mu_{0}H_{W}^{(b)}$ &$\mu_{0}H_{S}^{(a)}$ & $\mu_{0}H_{S}^{(b)}$  & $\mu_{0}H_{S}^{(c)}$    & $v_{W}$  & $v_{sat}^{*}$ & $D_{s}^{vsat}$  \\ \hline
                 &                 & [mT]               &   [mT]                  & [mT]                  &  [mT]                &    [mT]                        &  [m/s]   &   [m/s]  & [pJ/m]          \\ \hline

Pt/Co(0.5)/Pt         &     0.66        &   7.5              &11.5                       &  18.9                 &  29                   &     25              &           &               &                      \\

Pt/Co(0.8)/Ta(0.16)/Pt &  1.04         &  15.3             &19.7                    & 45.3                   &  58.15                  &    52.1                &  67.7                 &               \\

Pt/Co(0.8)/Ta(0.32)/Pt &  1.94      &  30.2              & 33.4                    & 89.5                &   99                 &  93.74              &  102.9                 &                  \\

Pt/Co(0.8)/Au         & 4.82       &  93.3              & 95.2                     & 234.5                     &    239.3                 &     237.2            &   259     & 250           &       -1.00 [-1.03]          \\

Pt/Gd$_{77}$Co$_{23}$(4)/Ta         & 5.21        & 42.4               &  43.2                     & 106.6                   &  108.5                  &      107.3            &   252     &  275          &            -0.8  [-0.78]    \\

\hline

    \end{tabular}
    \end{table*}

\section{Domain wall dynamics under out-of-plane magnetic field: 1D model and 2D micromagnetic simulations}
In our previous work \cite{Pham2016,Krizakova2019} we showed that for sufficiently large DMI strength, the DW speed versus $B_{z}$ does not decrease after the Walker field $\mu_0 H_{W}$, but a plateau with constant velocity $v_{W}$ is observed, the length of the plateau being proportional to $D/M_{s}$. 
Since the Walker velocity $v_{W}$ can be expressed analytically as a function of $D$, its measurement allows the DMI strength to be obtained (Eq. \ref{eq:vW}). 
We have also shown that the end of the velocity plateau is expected to occur at the Slonczewski field $\mu_0 H_{S}$ \textit{i.e.} at the field for which, in the negative mobility region after the Walker field, a one-dimensional (1D) DW reaches the minimum velocity \cite{Krizakova2019}.  In Krizakova \textit{et al.} \cite{Krizakova2019} we have calculated both $\mu_0 H_{W}$ and $\mu_0 H_{S}$ in the case where the field $\mu_0H_D=\dfrac{\pi}{2}\mu_0H_{\mathrm{DMI}}$ is much larger than the DW demagnetizing field $H_{DW} $. Since this is not the case for Pt/Co/Pt and the Pt/Co/Ta/Pt samples studied here (see Table II), we extend our treatment to the general case, that includes small DMI values.
By comparing the experimental curves with the results of 1D model and micromagnetic simulations, we show that care should be taken to extract the DMI strength from the experimentally observed plateau velocity in the case where the DMI field is expected to be comparable to the DW demagnetizing field.

\subsection{Determination of $H_{W}$ and $H_{S}$: $q-\phi$ collective model}
Within a one-dimensional formalism, a domain wall can be  described with two collective coordinates: the DW position, $q$, and the internal angle of the magnetization within the DW, $\phi$. By expressing the magnetization in spherical coordinates, and considering a Bloch-like profile as ansatz, the Euler-Lagrange equations are given by \cite{Thiaville2012}:
\begin{align}
  \alpha  \frac{\dot{q}}{\Delta}+\dot{\phi}&=\gamma_0 H_z  \label{Eq2} \\
      \frac{\dot{q}}{\Delta}-\alpha\dot{\phi} &=\gamma_0(H_D\cdot \sin{\phi}-  H_{\mathrm{DW}} \sin{\phi}\cos{\phi})  \label{Eq3}
\end{align}
where  $\mu_0H_{DW} =\dfrac{2K_{DW}}{M_s}$ , with $K_{DW}=N_x\dfrac{\mu_0M_s^2}{2}$ the domain wall magnetic shape anisotropy and $N_x \approx \dfrac{t\ln{2}}{\pi\Delta}$ the demagnetizing factor of a N\'{e}el DW.

The DW moves steadily, $\dot{\phi}$=0,  until the applied field reaches the Walker field given by \cite{Thiaville2012}:
\begin{equation}
   H_W=\alpha \sin{\phi_W}(H_D-H_{DW} \cos{\phi_W})
   \label{Eq4}
\end{equation}
where $\cos{\phi_W}=\dfrac{\delta-\sqrt{\delta^2+8}}{4}$, with $\delta=\dfrac{H_D}{H_{DW} }$. The domain wall velocity at the Walker field (\textit{i.e.} the Walker velocity), is given by:
\begin{equation}
    v_W=\dfrac{\gamma_0\Delta}{\alpha}H_W=\gamma_0\Delta H_D\Big(\sin{\phi_W}-\dfrac{\sin{2\phi_W}}{2\delta}\Big)
    \label{Eq5}
\end{equation}
with $\gamma_0=\mu_0 \gamma$ where $\gamma$ is the gyromagnetic ratio.  Note that, as  previously reported \cite{Pham2016,Krizakova2019},  when $H_D\gg H_{DW} $, $H_W \approx\alpha H_D$ and Equation \ref{Eq5} reduces to:

\begin{equation}
    v_W\approx\dfrac{\pi }{2}\gamma\dfrac{D}{M_s}=\dfrac{\pi }{2}\gamma\dfrac{D_s}{M_s t}
    \label{eq:vW}
\end{equation}
This is the approximated expression that we used in Refs. \cite{Pham2016} and \cite{Chaves2019} to extract the DMI strength from the experimentally observed plateau velocity after the Walker field for various Pt/Co/MOx trilayers.

Above the Walker field,  $\phi$ can no longer keep a constant value and the DW magnetic moment starts precessing. In the first stage of the precessional regime, the differential mobility is negative so that a straight domain wall is unstable \cite{Slonczewski1972a}. The Slonczewski field $\mu_0H_S$, where the negative mobility regime stops and the DW velocity  reaches a  minimum, can be obtained in the general case, when both $H_{DW}$ and $H_{D}$ are finite, using a semi-analytical method. The complete derivation  is detailed in Appendix A.  However, when either $H_{DW}\ll H_{D}$, or $H_{D}\ll H_{DW}$, $H_S$ and $v_S$ (the velocity at $H_S$)   are well approximated by the following expressions:
\begin{align}
     H_S&=H_W\cdot \Big(\dfrac{1+ \alpha^2}{\alpha \sqrt{\alpha^2+2}}\Big) \label{Eq7} \\
     v_S&=v_W\cdot\Big(\dfrac{\alpha \sqrt{\alpha^2+2}}{1+ \alpha^2}\Big) \label{Eq8}
\end{align}
with $H_W$ and $v_W$ calculated from Eq. \ref{Eq4} and \ref{Eq5}.

Beyond the Slonczewski field, the DW velocity reaches again the 1D regime, with an asymptotic mobility $m=\gamma_{0}\Delta/(\alpha+\alpha^{-1})$.
In Table II we report, together with the value of $\delta=H_{D}/H_{DW} $,  the values of $\mu_{0}H_W$ and $\mu_{0}H_S$ calculated for the Pt/Co/Pt, Pt/Co/Ta/Pt, Pt/Co/Au and Pt/Gd$_{77}$Co$_{23}$/Pt samples, using the magnetic parameters reported in Table I and setting $\alpha$=0.3. For $\mu_{0}H_W$ we report both the approximated value ($\mu_{0}H_{W}=\alpha \mu_0H_{D}$) valid when $\delta \gg 1$ (a) and the exact value  obtained from  Eq. \ref{Eq4} (b). For $\mu_{0}H_S$ we report the approximated value obtained from Eq. \ref{Eq7} with approximated $\mu_{0}H_W$ (a), the value obtained from Eq. 7 with the exact $\mu_{0}H_W$ (b) and the exact value obtained from the semi-analytical approach described in Appendix A (c).

The difference between approximated and exact values of $\mu_{0}H_W$ and $\mu_{0}H_S$ decreases as $\delta$ increases. For Pt/Co/Pt and Pt/Co/Ta/Pt, where $\delta \approx$ 1, the difference between the two values can be as large as 35\%  and the approximation is not justified. On the other hand for Pt/Co/Au and Pt/Gd$_{77}$Co$_{23}$/Ta, where  $\delta \gg$ 1, the difference between the two values is of the order of 2\%  and  $\mu_{0}H_W$ and $\mu_{0}H_S$ can be easily obtained using the approximate expressions relying  on the measured $H_{DMI}$ field.

The domain wall speeds obtained solving Eqs. \ref{Eq2} and \ref{Eq3} for several magnetic field values before and after the Walker field are shown in Fig. \ref{fig:Figure2}, where they are compared with the measured velocities and with the 2D micromagnetic simulations described below.

\subsection{Micromagnetic simulations}

\begin{figure*}
  \begin{center}
    \includegraphics[width=16cm]{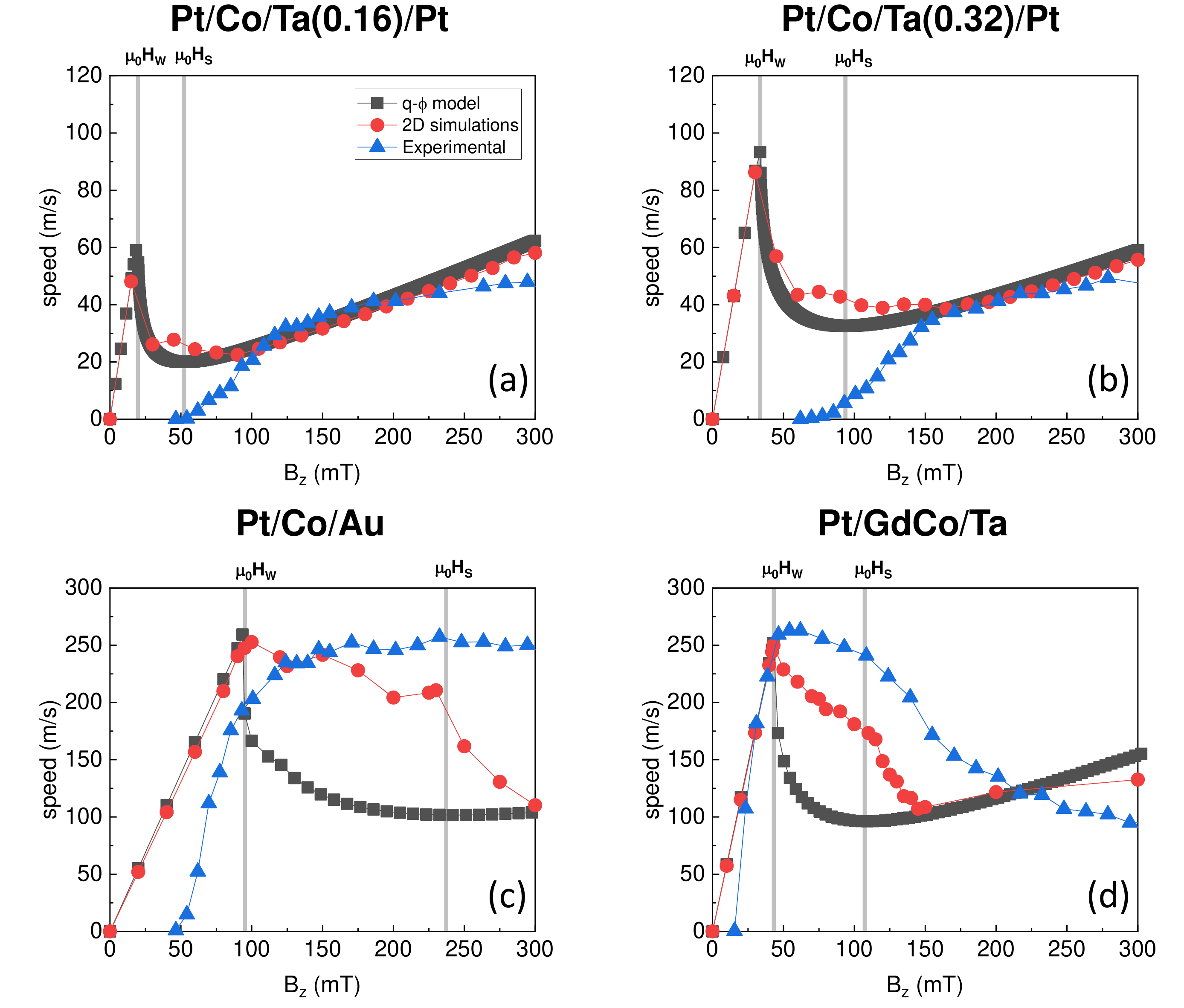}
  \end{center}
\caption{\label{fig:Figure2}\textbf{Domain wall speed versus B$_z$ field}: Domain wall velocity versus out-of-plane magnetic field $B_{z}$: (a) Pt/Co/Ta(0.16)/Pt;(b) Pt/Co/Ta(0.32)/Pt; (c) Pt/Co/Au; (d) Pt/Gd$_{77}$Co$_{23}$/Ta. The gray lines represent the exact values of the Walker field $\mu_{0}H^{(b)}_{W}$ and the Slonczewski field, $\mu_{0}H^{(c)}_{S}$, according to the 1D model. }
\end{figure*}

In order to include the 2D effects expected for domain walls in continuous thin films in the presence of DMI, micromagnetic simulations were carried out for the Pt/Co/Ta(0.16 and 0.32)/Pt, Pt/Co/Au and Pt/Gd$_{77}$Co$_{23}$/Ta samples. All the dynamic micromagnetic simulations are realized using \mbox{Mumax3} software \cite{Vansteenkiste2014} which solves the Landau-Lifshitz-Gilbert equation using a  finite-differences discretization. The material parameters reported in Table I, and the exchange stiffness values  reported above, were used for the simulations.  The  Gilbert damping parameter was initially chosen to be $\alpha$ = 0.3 for all the samples, and then slightly tuned to better fit the experimental data. The value $\gamma_0= 2.21 \times 10^5$~m/(A.s) was used throughout. For each simulation, a N\'{e}el domain wall set into 1$\mu$m wide strips (allowing 2D effects to occur) is  displaced by the action of a magnetic field normal to the plane and  is computed in a~$(1 \times 1)~\mu\text{m}^2$ moving-frame window  so as to keep the domain wall in its center. A 8~ns long field pulse is applied instantaneously at time $t$=0.  The cell size  is chosen to be (2 $\times$ 2)~{nm}$^2$ as it is sufficiently accurate with respect to the exchange length of around 10~nm in the cases examined here. The simulations are performed at 0~K in a defect-free sample. The DW velocities are calculated from the DW displacements driven by the applied field pulse, after inertia effects have disappeared. These  are shown in Fig.~\ref{fig:Figure2} where they are compared to those obtained with the $q-\phi$  model. More details are given in the Supplementary Information file \cite{SupplMat}. 

For all the samples, a good agreement between the results of 1D model and 2D simulation is found for the steady flow regime up to the Walker field. For the Pt/Co/Ta(0.16 and 0.32)/Pt stacks, the speed curves obtained with the two approaches are practically indistinguishable, as expected for weak DMI values \cite{Krizakova2019}. A slight difference between the Slonczewski fields found with the 1D model and the 2D micromagnetic simulations is seen for Pt/Co/Ta(3.2)/Pt,  where the DMI strength is slightly larger. On the other hand, for Pt/Co/Au and Pt/Gd$_{77}$Co$_{23}$/Pt stacks, where the interfacial DMI is much stronger, the 2D simulations predict a plateau of velocity after the Walker field and, as suggested in our previous work \cite{Krizakova2019} but not calculated before, the field for which the DW velocity starts dropping to join the precessional flow regime is in reasonable agreement with  the Slonczewski field $\mu_{0}H_{S}$ obtained with the 1D model.

\section{Domain wall dynamics under out-of-plane magnetic field: experiments and discussion }

Let us now compare the measured DW velocities versus out-of-plane field  with the predictions of the 1D model and the 2D micromagnetic simulations.

The domain wall velocities versus out-of-plane field $B_{z}$ for the Pt/Co/Ta(0.16 and 0.32)/Pt, Pt/Co/Au and Pt/Gd$_{77}$Co$_{23}$/Ta are reported in  Fig.~\ref{fig:Figure2}. The Pt/Co/Pt velocity curve (not shown here, and similar to that reported in \cite{Pham2016}) has the features expected for a domain wall in the presence of large disorder; only the creep  regime is visible. The depinning field is much larger than the Walker and the Slonczewski fields, so that the steady flow regime is hidden by the creep regime and the precessional flow is not reached for the largest fields. The small DMI has no visible effect on the DW dynamics driven by a $B_{z}$ field.

The DW velocities for the two Pt/Co/Ta/Pt samples have similar variations with  $B_{z}$: a large depinning field of  $\approx $100~mT and a slow increase of the DW speed, averaging at around 40-50 m/s,  for larger magnetic fields. Without the results of the 1D model and the 2D simulations, we might be tempted to associate this trend to the saturation of the DW velocity after the Walker field, due to 2D effects induced by the presence of DMI. However, the 1D calculation using the DMI value obtained from the experimental $\mu_0 H_{DMI}$ field predicts a drop of the DW velocity immediately after the Walker field. The Walker field could not be observed, because it occurs for magnetic field values (resp. $\approx$ 20~mT and $\approx$ 40~mT) well below the measured depinning field. 
As a consequence, for these two samples, where the DMI field is comparable to the DW demagnetizing field ($\delta\approx$1 in Table II), Eq. \ref{eq:vW} should not be used to extract the DMI value from the apparent saturation DW velocity, since this does not correspond with the Walker velocity.  Instead, the calculations, carried out using a damping constant $\alpha$=0.25 for the two samples, show that the slow increase of the DW velocity observed for large fields is most probably associated to the constant mobility expected in the precessional regime.

The situation is different for the Pt/Co/Au and Pt/Gd$_{77}$Co$_{23}$/Ta samples, where the velocity versus in-plane field measurements indicate the presence of a large interfacial DMI ($\delta\gg$1). For Pt/Co/Au, the calculated Walker field (around 90~mT) is much larger than that of the previous samples, but it is situated within the thermally activated regime, so that the steady flow regime is not observable. However, the measured plateau velocity fits remarkably well with the calculated Walker velocity obtained using Eq. 5 and the DMI value extracted from the  measured $\mu_0 H_{DMI}$ field (Table II) and with the results 2D simulations. This result validates the method consisting in measuring the Walker velocity  (or the plateau DW velocity) to extract the DMI value, and also confirms the validity of the exchange stiffness A=16~pJ/m used to extract the DMI from the DMI field. 
Note however that, in contrast with the 2D simulations, the experimental DW velocity conserves the Walker velocity beyond the calculated Slonczewski field and up to the highest measured fields. We suggest that  this  might be related to a dissipation mechanism associated to disorder \cite{Voto2016},  in particular to the likely inhomogeneity of the PMA at the scale of the grain size, which may modify the DW dynamics in a way that cannot be taken into account easily by theoretical models. This has been confirmed qualitatively by our preliminary micromagnetic simulations including disorder.

For Pt/Gd$_{77}$Co$_{23}$(4)/Ta the velocity curve is different from those observed for the Pt/Co/M stacks. The much weaker value of the depinning field, smaller than any of the depinning fields observed in our samples so far, is most probably due to the large DW width associated to the low effective anisotropy and low exchange stiffness.  This allows us to observe the end of the steady flow regime and the Walker field, whose value corresponds well with that predicted by the 1D model. The slope of the steady state regime also has a good correspondence with the theory, which validates the value used for the Gilbert damping parameter. The velocity stays almost constant for fields above $H_W$, before decreasing rapidly above 100~mT. The change of slope of the velocity curve occurs for a field close to the predicted Slonczewski field, and the precessional flow regime is reached around 200 mT. After the Walker field, the measured DW velocity decreases less rapidly than expected from the 2D simulations. This might be associated to the presence of disorder, as in the case of Pt/Co/Au.

To summarize,  for Pt/Co/Au and Pt/Gd$_{77}$Co$_{23}$/Ta, for which $H_{D}\gg H_{DW} $, the observed plateau velocity corresponds with the Walker velocity $v_{W}$.  For these two samples, the measurement of the plateau DW velocity provides a robust means to obtain the strength of the interfacial DMI since, from Eq. \ref{eq:vW}:
\begin{equation}
\label{eq:Ds}
D_{s} \approx \frac{2}{\pi\gamma}M_{s}t~v_{W}
\end{equation}

The DMI values obtained using this expression are reported in Table II, where they are compared with the values obtained from the measured $\mu_0 H_{DMI}$ field. This method is particularly appealing since it only relies on easily measurable parameters and does not require the knowledge of the exchange stiffness, whose exact value is rarely measured for ultrathin films. The adjustment of the DMI value obtained using the $\mu_0 H_{DMI}$ field to that obtained from the plateau velocity provides a way to measure the exchange stiffness of these thin films. In the case of Pt/Co/Au, A=16 pJ/m obtained from previous studies for stacks fabricated with the same method provides a good agreement between the two DMI values. A smaller value of the exchange stiffness is obtained for Pt/Gd$_{77}$Co$_{23}$/Ta, in agreement with data reported in the literature for ferrimagnetic alloys \cite{Mansuripur1986}.

This method is not applicable to the two Pt/Co/Ta/Pt stacks, where $\delta = H_{D}/H_{DW}\approx 1 $ and the observed DW velocity behaviour for large field cannot be associated to the velocity plateau produced by the 2D effects induced by the DMI.  

We may now wonder how, without an alternative measurement of the DMI strength (as it was done here by measuring the DW velocity versus $B_z$ under a static $B_x$), we may analyze a general DW velocity curve showing a large depinning field and velocities apparently saturating at large fields, in order to acquire information on the strength of $\delta$, and therefore on the applicability of our method. 
We assume that the sample has been characterized magnetically, so that $M_\mathrm{s}$, $\Delta$, $H_\mathrm{DW}$ and
$\alpha$ are known. A systematic approach would be to numerically solve the 1D model (Eq.~2 and~3) for increasing values of $H_\mathrm{D}$ to see which $v(H_z)$ curve fits bets the data, but this may be too cumbersome. In order to get more directly some estimates, one may first assume that the observed plateau velocity ($v^*$) corresponds to the Walker velocity. One can then extract $D^*$ using Eq. 6, and compute the DMI field from Eq.~1, leading to the evaluation of $\delta = (\pi/2) H_\mathrm{DMI}/H_\mathrm{DW}$. If $\delta \gg 1$ the assumption was correct. An additional check is to compare the minimum value of the field for which the experimental DW velocity saturates ($H^*$) with the derived $H_S$, field above which the predictions of the 1D model become valid again. If $H^*$ is smaller than $H_S$, then the plateau velocity corresponds with the Walker velocity and the $D^*$ value is correct. If $H^*$ is larger than $H_S$, the plateau velocity does not correspond with the Walker velocity, but it is likely to be close to the Slonczewski velocity, especially when the damping is small. One may therefore tune $D$ in order to fit $v^*$ with the Slonczweski velocity predicted by the 1D model (Eq. 8).

\section{Conclusion}

In conclusion, the domain wall dynamics driven by an out-of-plane magnetic field well beyond the Walker field was measured for several magnetic trilayers with different strengths of the interfacial Dzyaloshinskii-Moriya interaction.  Using the method consisting in measuring the asymmetric expansion of a magnetic domain in the presence of an in-plane field, we have obtained the values of the $\mu_0 H_{DMI}$ fields from which we have evaluated the  DMI constant, using exchange parameters $A$ derived in previous works. The DW velocity curves versus $B_{z}$ field were then  examined in the light of a 1D analytical model, by extending the one presented in our previous works to the case of samples where the DMI field is comparable with the DW demagnetizing field. The exact semi-analytical solution for the Slonczewski field was obtained. Experimentally, the Walker velocity, which can be maintained after the Walker field in systems with large DMI, can  be related to the strength of the DMI, and its measurement provides a robust approach to obtain the DMI constant and the exchange stiffness. By comparing the measured DW speed curves with the expectations of the 1D model and of the 2D simulations, we show that care should be taken when using this approach with samples with a weak DMI field. 

\section{Acknowledgements }
\begin{acknowledgements}
We acknowledge the support of the Agence Nationale de la Recherche (projects ANR-17-CE24-0025 (TOPSKY)). A.F., S.P. and J.V. have been supported by the the DARPA TEE program through Grant No. MIPR HR0011831554. J.P.G. acknowledges the European Union’s Horizon 2020 research and innovation program under Marie Sklodowska-Curie Grant Agreement No. 754303 and the support from the Laboratoire d'excellence LANEF in Grenoble (ANR-10-LABX-0051). B. Fernandez, Ph. David, D. Lepoittevin and E. Mossang are acknowledged for their technical help.
\end{acknowledgements}

\section{Appendix A: Calculation of the Slonczewski field}

Combining the two equations (\ref{Eq2}) and (\ref{Eq3}) of the 1D model, including the presence of an external in-plane magnetic field $\mu_0H_x$, the autonomous
evolution of the angle $\phi$ with time is obtained, from which the precession period $T$ is found
\begin{eqnarray}
T &=& \frac{1+\alpha^2}{\gamma_0} \int_0^{2 \pi} \frac{d \phi}{H_z-\alpha \sin\phi \left(H_D + \dfrac{\pi}{2}H_x-H_{DW} \cos\phi\right)} \nonumber \\
&\equiv& \dfrac{1+\alpha^2}{ \gamma_0 \alpha H_D} I_1 \left[h_z, h_x\right]
\label{EqA1}
\end{eqnarray}
where
\begin{equation}
   I_1 \left[h_z, h_x\right]= \int_0^{2 \pi} \dfrac{d \phi}{h_z-\sin\phi \Big(1+h_x - \dfrac{\cos\phi}{\delta} \Big)} \nonumber
\end{equation}
is a function of the variables $h_z=\dfrac{H_z}{\alpha H_D}$ and $h_x=\dfrac{H_x}{H_{\mathrm{DMI}}}$, with $\delta$ a parameter.  
Note that Eq.~\ref{EqA1} was already considered in Ref.~\cite{Kim2019} when studying the domain wall precessional motion under DMI and in-plane field.
From Eq.~(\ref{Eq2}), by taking the average over one precession period, the average DW velocity reads
\begin{equation}
\langle \dot{q} \rangle= \frac{\gamma_0 \Delta}{\alpha} H_z - \frac{\Delta}{\alpha} \frac{2 \pi}{T}.
\label{EqA2}
\end{equation}
From the expression of $I_1[h_z.h_x]$, it becomes clear that (when $\mu_0H_x=0$, i.e. $h_x=0$) the curve $\langle \dot{q} \rangle$(H$_z$)  has the same shape when $H_D \gg H_{DW}$, compared to the standard case $H_D \ll H_{DW}$.
This is the reason behind the relations (\ref{Eq7}) of $H_S$ to $H_W$, and (\ref{Eq8}) of  $v_S$ to $v_W$.
But in the general case, the shape of the curve is different.

In order to obtain the exact Slonczewski field, the minimum of $\langle \dot{q} \rangle$ versus $H_z$ has to be found.
From Eq.~(\ref{EqA2}), it is given by the solution of
\begin{equation}
\dfrac{1+\alpha^2}{2\pi} =\dfrac{I_2 \left[h_z,h_x \right]}{I_1^2 \left[h_z,h_x \right]},
\label{EqA3}
\end{equation}
where a similar integral $I_2$ appears, again as a function of $h_z$ with $\delta$ as parameter
\begin{equation}
I_2[h_z,h_x] = \int_0^{2 \pi} \dfrac{d \phi}{\Big[h_z-\sin\phi \Big(1+h_x- \dfrac{\cos\phi}{\delta} \Big)\Big]^2}. \nonumber
\end{equation}

The integrals $I_1$ and $I_2$ are easily evaluated by numerical integration, for all values of their argument $h_z$ (with $H_z > H_W$) and parameter $\delta$. Here we consider the case $\mu_0H_x=0$, i.e. $h_x=0$. In Appendix B, we will consider a more complex situation where $h_z$ and $h_x$ are non-zero. 

Taking into account the value of $\alpha$, the value of $H_S/(\alpha H_D)$ is found by numerically solving Eq.~(\ref{EqA3}), and the Slonczewski velocity $v_S$ follows by inserting $H_z=H_S$ into Eq.~(\ref{EqA2}).
\begin{figure}[t]
\includegraphics[width=8cm]{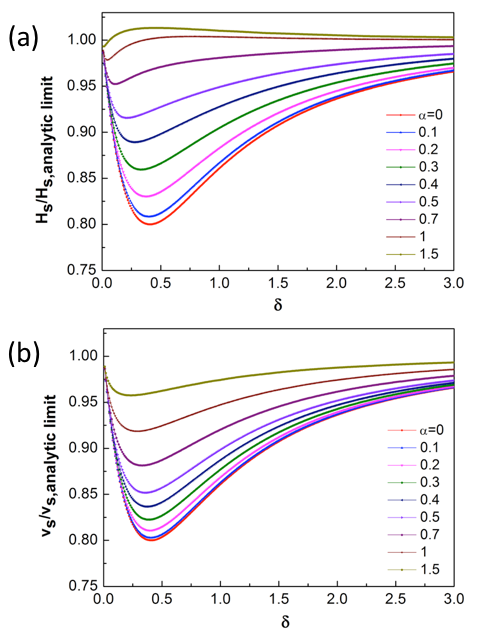}
\caption{
\label{fig:Fig-A1}
\textbf{Semi-analytical Slonczewski field $H_S$ (a) and velocity $v_S$ (b) compared to the limiting
analytical expressions Eq.~(\ref{Eq7}) resp. (\ref{Eq8}).}
The ratio of the two expressions is plotted versus  $\delta=H_D/H_{DW}$  for different values of the damping parameter $\alpha$.
The curves for $\alpha=0$ are analytic (see text).}
\end{figure}
In order to get an idea of the correction involved in solving exactly the $q-\phi$ model in the presence of both DMI and domain wall magnetic shape anisotropy, a plot of $H_S$ normalized to the limiting analytical value (Eq.~(\ref{Eq7})) versus $\delta$ and for different values of $\alpha$, is provided in Fig.~\ref{fig:Fig-A1}(a).
One sees that $H_S$ is mostly smaller than the analytic limiting value, the maximum difference strongly depending on the value
of $\alpha$.
For large damping ($\alpha \geq 1$), the exact solution can be slightly larger than the limiting analytical expression.
A similar plot is shown in Fig.~\ref{fig:Fig-A1}(b), for the Slonczewski velocity, this time compared to the limiting
expression Eq.~(\ref{Eq8}).
The curves have the same overall shape, but different numerical values.
One notices that the exact solution gives velocities always smaller than the analytical limit.

The above plots contain also the solution of the problem in the limit $\alpha \ll 1$, which is analytic for every value of $\delta$.
This last calculation leads to
\begin{equation}
H_S= H_{DW} \sqrt{\frac{1+(2 H_D/H_{DW})^2}{8}}
\end{equation}
and
\begin{equation}
v_S= \alpha \gamma_0 \Delta H_{DW} \sqrt{\frac{1+(2 H_D/H_{DW})^2}{2}}.
\end{equation}
These relations are plotted in Fig.~\ref{fig:Fig-A1}(a), resp. Fig.~\ref{fig:Fig-A1}(b), where they are  normalized to Eq. \ref{Eq7}, resp. Eq. \ref{Eq8}.

\section{Appendix B: In-plane field giving a minimum velocity in the precessional regime}

In the precessional regime, the average DW velocity over one period is given by Eq. \ref{EqA2}. From this equation, the minimum average velocity for a fixed $H_z$,  occurs when the right term is maximum (it is positive as $H_z>H_W$) i.e.,  when the period $T$ is minimum. The minimisation of $I_1$ with respect $h_x$ leads to 
\begin{equation}
 \frac{\partial I_1}{\partial h_x} = 2(1+h_x) \int_0^{\pi}{\dfrac{\sin^2 \phi d\phi}{\left(h_z+\dfrac{\sin \phi \cos \phi}{\delta}\right)^2-(1+h_x)^2\sin^2\phi}}
\end{equation}
which it is equal to zero when $h_x=-1$. Therefore, the average velocity, Eq. \ref{EqA2} will be minimum when $H_x=-H_{\mathrm{DMI}}$. The result is rigorous if, for all values of $H_x$ considered, one has $H_z > H_\mathrm{S} \left[ H_x \right]$: no 2D instability occurs and the 1D precessional regime is observed.

Note that $H_\mathrm{S} \left[H_x \right]$ is obtained by solving Eq.\ref{EqA3}, with a non-zero $h_x$.

\begin{figure}[t]
\includegraphics[width=8cm]{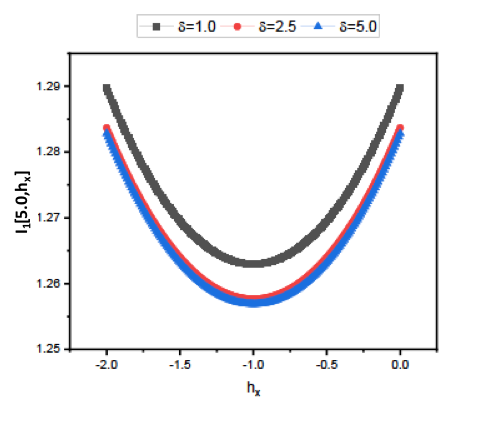}
\caption{
\label{fig:Fig-AppB1}
\textbf{Numerical calculation of $\mathbf{I_1 [h_z,h_x]}$ versus $\mathbf{h_x}$ for different values of $\mathbf{\delta=1.0, 2.5, 5.0}$, with  $h_z=5.0$}}   
\end{figure}

Figure \ref{fig:Fig-AppB1} shows the numerical integration  $I_1\left[h_z, h_x\right]$ \textit{vs} $h_x$ for $h_z=5.0$ (above the Slonczewski field), for different values of $\delta$. A parabolic curve centered in $h_x=-1$, with a curvature depending on the values of $h_z$ and $\delta$ is observed.  As expected from the above calculation, regardless of the value of $h_z$ and $\delta$,  the minimum occurs when $h_x=-1$, or equivalently when  $H_x=-H_{\mathrm{DMI}}$. 
This conclusion was already reached in Ref.~\cite{Kim2019}.

\section{References}
%
\end{document}